%% file: na59-l4.tex
\documentclass[aps,pra,twocolumn,showpacs,showkeys,superscriptaddress]{revtex4}

\usepackage{graphicx}
\usepackage{dcolumn}
\usepackage{bm}

\begin{document}


\title{Linear to Circular Polarisation Conversion using Birefringent 
       Properties of Aligned Crystals for Multi-GeV Photons}



\input{na59-l4-authors.tex}

\noaffiliation

\date{\today}

\begin{abstract}
We present the first experimental results on the use of a thick aligned Si
crystal acting as a {\it quarter wave plate} to induce a degree of
circular polarisation in a high energy linearly polarised photon beam. The
linearly polarised photon beam is produced from coherent bremsstrahlung
radiation by 178\,GeV unpolarised electrons incident on an aligned Si
crystal, acting as a {\it radiator}. The linear polarisation of the photon
beam is characterised by measuring the asymmetry in e$^+$e$^-$ pair
production in a Ge crystal, for different crystal orientations. The Ge
crystal therefore acts as an {\it analyser}. The birefringence phenomenon,
which converts the linear polarisation to circular polarisation, is
observed by letting the linearly polarised photons beam pass through a
thick Si {\it quarter wave plate} crystal, and then measuring the
asymmetry in e$^+$e$^-$ pair production again for a selection of relative
angles between the crystallographic planes of the {\it radiator}, {\it
analyser} and {\it quarter wave plate}. The systematics of the difference
between the measured asymmetries with and without the {\it quarter wave
plate} are predicted by theory to reveal an evolution in the Stokes
parameters from which the appearance of a circularly polarised component
in the photon beam can be demonstrated. The measured magnitude of the
circularly polarised component was consistent with the theoretical
predictions, and therefore is in indication of the existence of the
birefringence effect.
\end{abstract}


\maketitle


\section{\label{sec:intro}Introduction}

The demand for high energy circularly polarised photon beams has increased
with the need to perform experiments to determine the gluon spin density
of the nucleon~\cite{compass,ric,bosted} from polarised photon-gluon
fusion, and polarised photo production of high transverse momentum
mesons~\cite{afanas}. A well known method to produce circularly polarised
photons is from the interaction of longitudinally polarised electrons with
crystalline media, where the emitted photons are circularly polarised due
to conservation of angular momentum~\cite{olsen}. Theoretical
calculations~\cite{nadz,armen} predict that the coherent
bremsstrahlung~(CB) and channelling radiation in crystals by
longitudinally polarised electrons are also circularly polarised, and can
be used to enhance the number of high energy circularly polarised photons.
Currently, the highest energy available for polarised electrons is only
45~GeV~\cite{slac1,slac2}. Therefore, the polarised electron beams are not
sufficiently energetic to give photons that will be in the kinematic
regime where $\gamma g\rightarrow c\bar{c}$ is well above the $\Lambda_c
D$ associate production, that is, for a centre of mass energy above
70~GeV.

We have been working toward testing the conjecture that it is possible to
produce circularly polarised photon beams in proton accelerators using the
extracted unpolarised high energy electron beams with energies of up to
250\,GeV (CERN) and 125\,GeV (FNAL)~\cite{propos}. These unpolarised
electron beams can produce linearly polarised photons via CB radiation in
an aligned single crystal. One can transform the initial linear
polarisation into circular polarisation by using the birefringent
properties of aligned crystals. The above mentioned method was first
proposed by Cabibbo and collaborators in the 1960's~\cite{cabibbo1}, and
later the numerical calculations were done in terms of coherent pair
production (CPP)~\cite{termisha} theory to obtain the optimal thicknesses
for various cubic crystals.

\begin{figure*}[ht]
\includegraphics[scale=0.42]{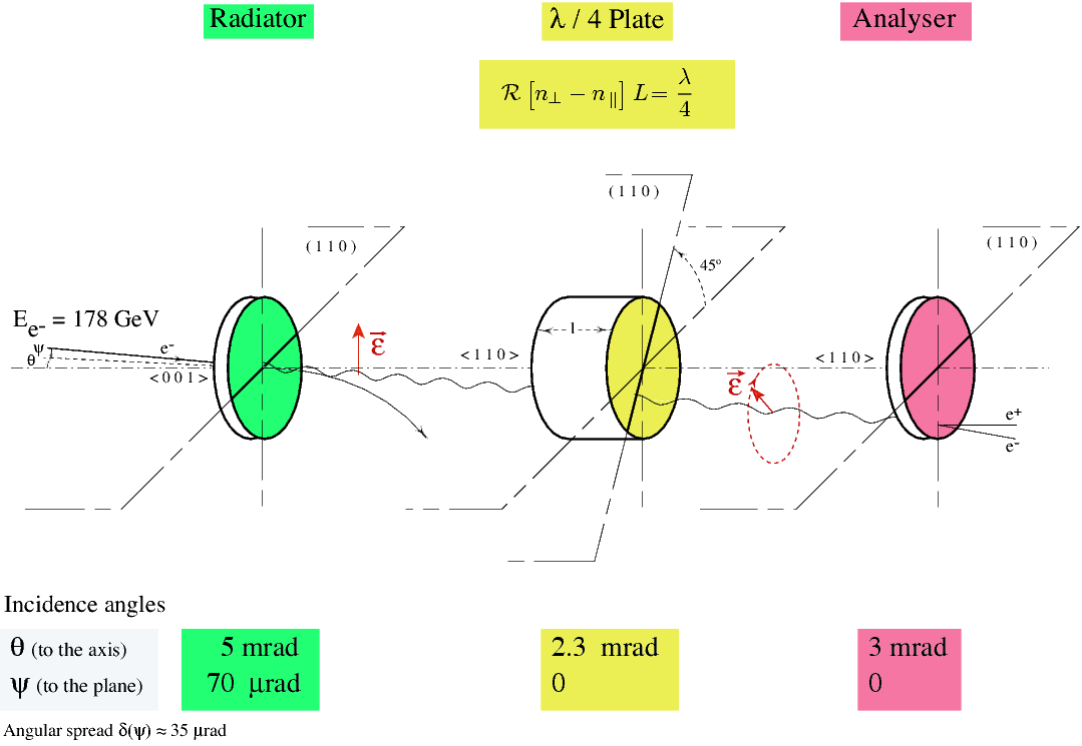}
\caption{\label{F:setup1} Three crystal scheme.}
\end{figure*}

One of the purposes of the NA59 collaboration was to investigate the
birefringent properties of aligned crystals, and test the feasibility of
producing high energy circularly polarised photon beams starting from the
unpolarised 178\,GeV electrons beams at CERN SPS~\cite{propos}. This
experiment used a consecutive arrangement of three aligned single
crystals: the first crystal acted as a radiator to produce a linearly
polarised photon beam, the second crystal acted as a quarter wave plate to
convert the linear polarisation into circular polarisation, and the last
crystal acted as an analyser to measure the change in the linear
polarisation of the photon beam. The three crystal scheme used is shown in
Fig.~\ref{F:setup1}.

The linearly polarised photon beam was produced by CB radiation from
electrons in an aligned Si\,$<$001$>$ single crystal. The orientation of
the crystal {\it radiator} was chosen in such a manner that the coherent
peak of the photon spectrum was situated between 70-100\,GeV. For the
conversion of the linear polarisation into circular polarisation an
aligned Si\,$<$110$>$ crystal was used. Finally an aligned Ge$<$110$>$
crystal was used as an {\it analyser} of the linear polarisation of photon
beam.

When high energy photons propagate through a medium, the main process by
which the photons are absorbed is e$^+$e$^-$ pair production~(PP). When
photons propagate through an aligned crystal at small incident angles with
respect to a crystal axis and/or a crystal plane, a coherent enhancement
of the PP is manifested (CPP). The cross section for the CPP process
depends on the direction of the linear polarisation of the photon beam
with respect to the plane containing the crystal axis and the photon
momentum (reaction plane) as shown in Fig.~\ref{F:setup1}. In the
experiment we used the configuration proposed in~\cite{cabibbo1}, where
the reaction plane coincides with the crystal plane $(110)$. Following the
description given in~\cite{cabibbo1}, one can represent the linear
polarisation of the photon beam as a superposition of two beams with
polarisation directions parallel and perpendicular to the reaction plane
containing the photon momentum {\boldmath $k$} and the crystallographic
axis. In this symmetric case, with respect to the plane containing the
photon momentum the birefringent effect is maximum and the photon
polarisation vector {\boldmath $\epsilon$} will be the combination of two
unit vectors, {\boldmath $t$} and {\boldmath $y$}, parallel and
perpendicular to the reaction plane, respectively:
\begin{equation}
\mbox{\boldmath $\epsilon$} = \mbox{\unboldmath$\epsilon_{\parallel}$} 
\mbox{\boldmath $t$} +
\mbox{\unboldmath$\epsilon_{\perp}$} \mbox{\boldmath $y$}.
\label{eq:no1}
\end{equation}

The components of the polarisation vector before and after the crystal of
thickness, $L$, are related by a $2 \times 2$
matrix~\cite{cabibbo1,zorzi}:

\begin{displaymath}
\left( \begin{array}{c}
\epsilon_{\parallel}(L)\\
\epsilon_{\perp}(L) 
\end{array} \right)
=
\left( \begin{array}{cc}
      \exp [in_{\parallel} E_\gamma L] & 0 \\
      0 & \exp [in_{\perp} E_\gamma L]
\end{array} \right)
\left( \begin{array}{cc}
      \epsilon_{\parallel}(0) \\
      \epsilon_{\perp}(0)
\end{array} \right)
\end{displaymath}
where $E_\gamma$={\boldmath $|k|$} ($\hbar=c=1$) is the energy of the
incident photon and the $n_\parallel$ and $n_\perp$ are complex quantities
analogous to the index of refraction. As was mentioned in ~\cite{cabibbo1}
"The diagonality is due to our choice of the basis vectors and to the
symmetry of our situation in respect to the plane". The general case, when
the photon momentum makes an small angle relative to the crystal plane was
considered in~\cite{maish1,maish2,strakh}. The imaginary part of the index
of refraction is connected with the photon absorption cross section, while
the real part can be derived from the imaginary part using dispersion
relations~\cite{cabibbo1}. The crystal can act as a {\it quarter wave
plate}, if the real part of the relative phases of the two components of
the waves parallel and perpendicular to the reaction plane is changed by
$\pi/2$ after transmission of the photon. Thus the crystal will be able to
transform the linear polarisation of the photon beam into circular
polarisation at the matching thickness:

\begin{equation}
L = \frac{2}{\pi}\frac{1}{E_\gamma\Re(n_{\perp} - n_{\parallel})}.
\label{eq:no2} 
\end{equation}

In this paper we determine the polarisation direction with respect to the
crystal radiator, and express this polarisation using Stokes parameters
(with the Landau convention). Referred to our geometry the parameter
$\eta_1$ describes the linear polarisation of the beam polarised in the
direction of $45^{\circ}$ to the reaction plane of the radiator, while the
parameter $\eta_3$ describes the linear polarisation in the direction
parallel or perpendicular to the reaction plane of the radiator. The
parameter $\eta_2$ describes the circular polarisation.  The total
polarisation is then written:
\begin{widetext}
\begin{equation}
P_{\hbox {linear}}=\sqrt{\eta _{1}^{2}+\eta _{3}^{2}},
\quad \; P_{\hbox {circular}}=\sqrt{\eta _{2}^{2}},
\quad \; P_{\hbox {total}}=\sqrt{P_{\hbox {linear}}^{2}+P_{\hbox
{circular}}^{2}} \quad .
\label{eq:pol-def}
\end{equation}
\end{widetext}

The photon beam intensity and Stokes parameters after the quarter wave
plate with the thickness $L$ can be derived from the following
formulae~\cite{baier0,baier}:
\begin{eqnarray}
N(L) &=& N(0)[\cosh aL+\eta _{1}\sinh aL]\exp (-WL),
\nonumber \\
\eta_1 (L) &=& \frac{\sinh aL + \eta_1 (0)\cosh aL}
              {\cosh aL +\eta_1 (0)\sinh aL},
\nonumber \\
\eta_2 (L) &=& -\frac{ \eta_3 (0)\sin bL - \eta_2 (0)\cos bL}
              {\cosh aL + \eta_1 (0)\sinh aL},
\nonumber \\
\eta_3 (L) &=& -\frac{\eta_3 (0)\cos bL + \eta_2 (0)\sin bL}
                {\cosh aL + \eta_1 (0)\sinh aL},
\label{eq:no4}
\end{eqnarray}
with
\begin{eqnarray}
a &=& E_\gamma\Im(n_{\perp} - n_{\parallel}) = \frac{1}{2}
(W_{\parallel}-W_{\perp}),
\nonumber \\  
b &=& E_\gamma\Re(n_{\perp} - n_{\parallel}),
\quad W = \frac{1}{2}(W_{\parallel}+W_{\perp}),
\label{eq:no5}
\end{eqnarray}
where $W_\parallel$ and $W_\perp$ are the pair production probabilities
per unit path length for photons polarised parallel or perpendicular to
the reaction plane, respectively.

As follows from equation~(\ref{eq:no4}), the component of the linear
polarisation in the direction of $45^{\circ}$ to the reaction plane of the
quarter wave crystal is transformed into circular
polarisation~\cite{baier}.  Therefore the {\it quarter wave plate} should
be rotated by $45^{\circ}$ with respect to the polar plane of the photon
beam to have the optimal transformation of the polarisation. In this case
the linear polarisation component $\eta_3$, which was defined as the one
parallel or perpendicular to the reaction plane of the radiator,
represents a component of the linear polarisation in the direction of
$45^{\circ}$ to the reaction plane of the quarter wave plate
equation~(\ref{eq:no4}).

\begin{figure*}[ht]
\includegraphics[scale=0.445]{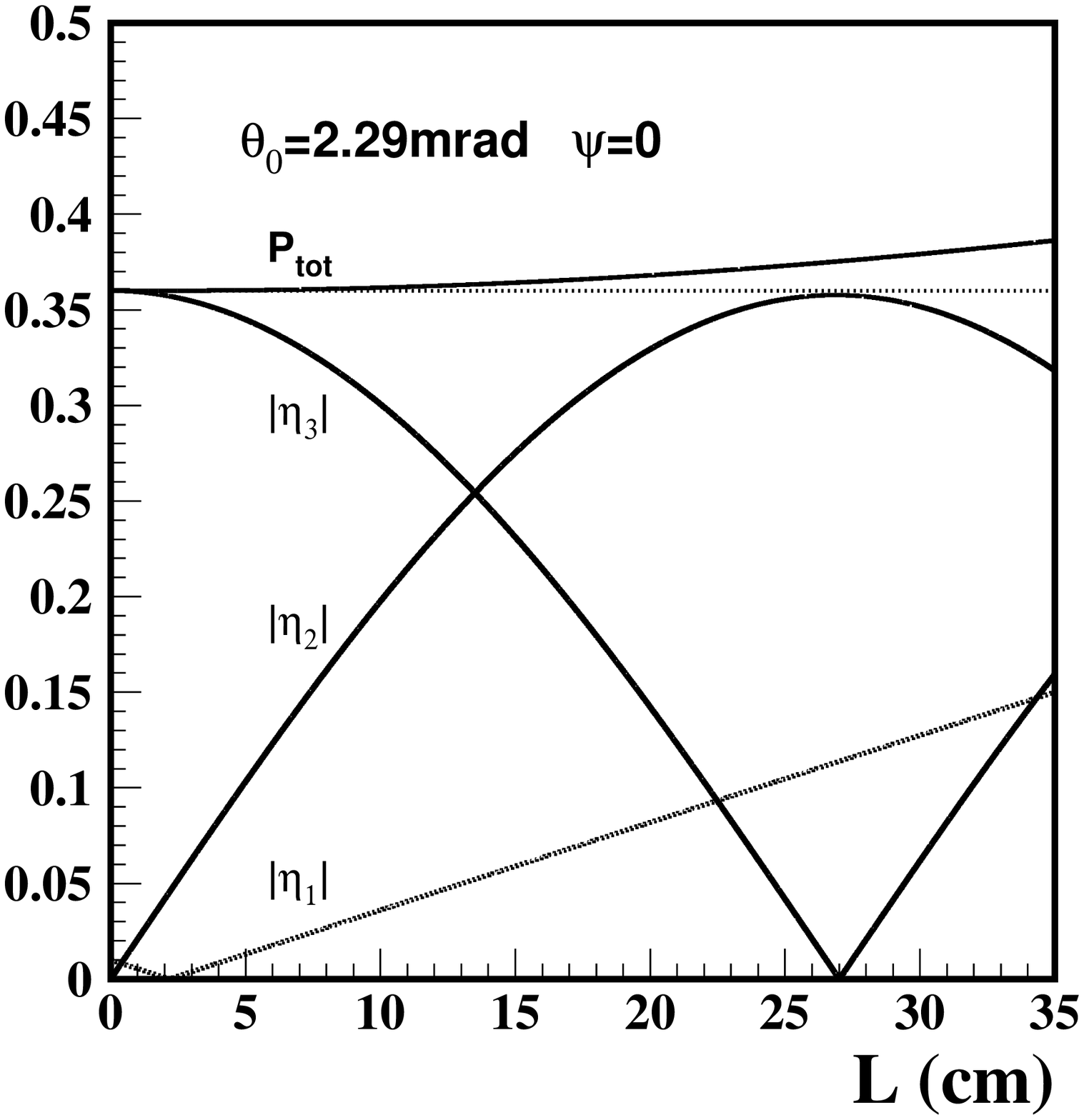}
\includegraphics[scale=0.445]{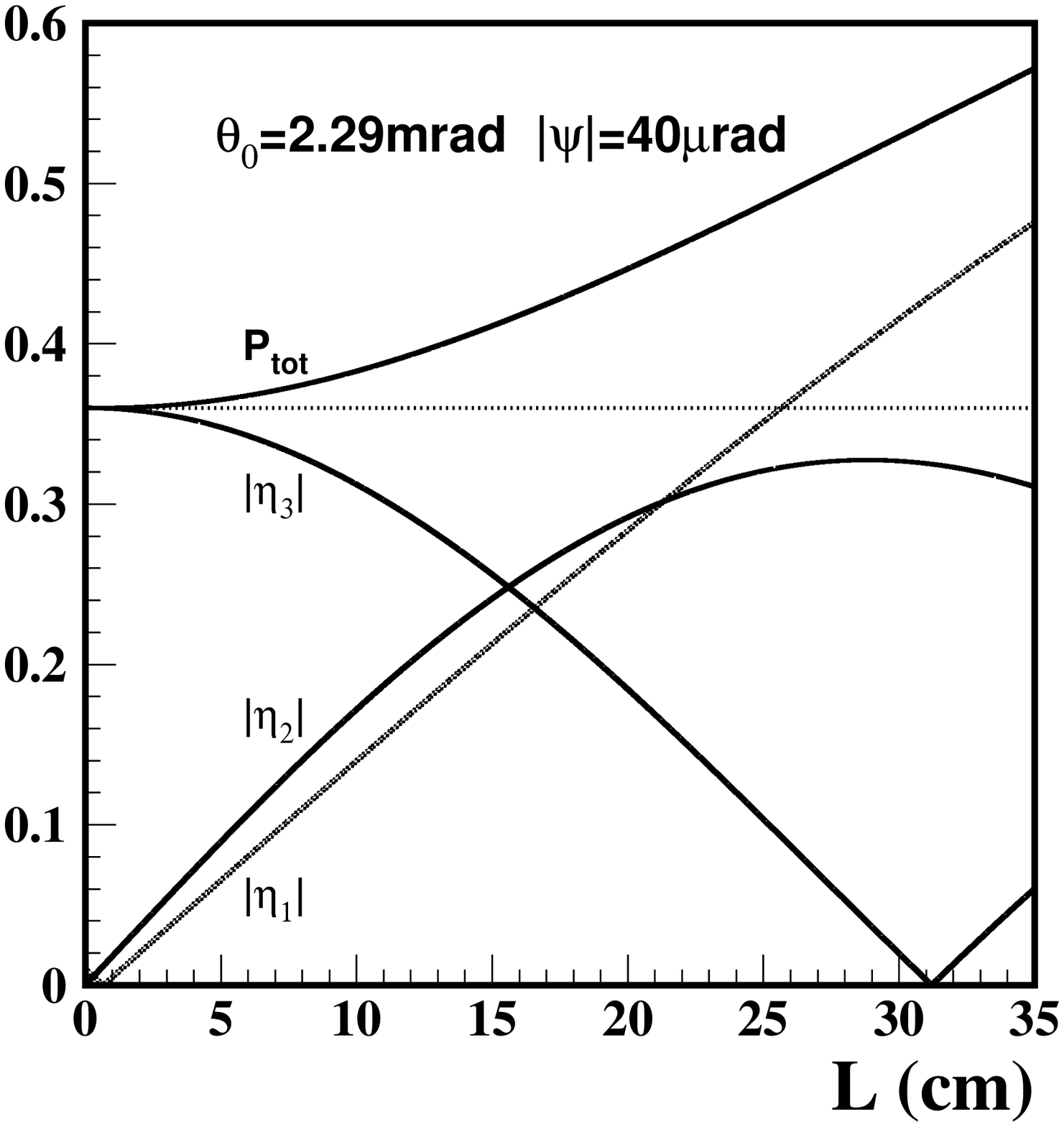}
\caption{\label{F:stokes} Absolute values of the Stokes parameters and 
   the total degree of polarisation for a Si crystal as a function of its 
   thickness $L$, for $E_\gamma$=100\,GeV linearly polarised  photons. The 
   left hand figure and the right hand figure are calculated using initial 
   values for the Stokes parameters described in the text. For these 
   conditions, the crystal also acts as polariser generating a $\eta_1$ 
   component.}
\end{figure*}

As follows from equation~(\ref{eq:no4}), the total polarisation of the
photon beam before and after the quarter wave crystal are connected by the
relation:
\begin{equation}
P_{total}^2(L) = 1 + \frac{P_{total}^2(0)-1} {(\cosh aL +
\eta_1 (0)\sinh aL)^2}.
\label{eq:no6}
\end{equation}

There is conservation of polarisation if the incident photon beam is
completely polarised. In a real experiment, the incident photon beam is
not completely polarised, and one must seek an alternative conserved
quantity. Further study of equation~(\ref{eq:no4}) reveals that the
quantity
\begin{equation}
K\equiv\frac{\eta_2^2(\ell) + \eta_3^2(\ell)}{1- \eta_1^2(\ell)}
\label{eq:no7}
\end{equation}
is constant and conserved when a photon beam penetrates the quarter wave
plate crystal~\cite{kononets} in exact symmetric orientation. This
relation holds for any penetration length, $\ell$, between $0 \leq \ell
\leq L$ except in the case when $\eta_2(0)=\eta_3(0) \equiv 0$ and
$\eta_1(0)=1$. It allows the determination of the resulting circular
polarisation of photon beam by measuring its linear polarisation before
and after the quarter wave plate. Taking into account the experimental
condition, i.e. the photon beam angular divergences, one can note that K
is conserved with $\approx$5$\%$ accuracy in the 80-110~GeV region as show
Monte-Carlo simulations.

Fig.~\ref{F:stokes} shows the expected dependence of the Stokes parameters
describing the photon polarisation as a function of the {\it quarter wave
plate} thickness, $\ell$, for the surviving photons from a beam of
100\,GeV. One can see 
1/4 wave action. that the initial total polarisation is not conserved in
the case of a partially polarised photon beam as expected from
equation~(\ref{eq:no6}), nevertheless, the relation \ref{eq:no6} still
holds.

These calculations were carried out assuming that the Stokes parameters
before the {\it quarter wave plate} had the following values:
$\eta_1$=0.01, $\eta_2$=0 and $\eta_3$=0.36. In Fig.~\ref{F:stokes}
(left), the photon beam makes an angle $\theta_0$=2.3\,mrad with respect
to the $<$110$>$ axis and lies in the $(110)$ plane ($\psi$=0), while in
Fig.~\ref{F:stokes} (right), the photon beam traverses the $(110)$ plane
at a small angle, $\psi$=$\pm$40\,$\mu$rad.

One can see the increase in the total polarisation, $P_{total}$, after the
{\it quarter wave plate} with respect to the initial total polarisation
(the straight line around 0.36). This difference comes from the fact that
the aligned {\it quarter wave plate} can also act as a polariser.
Therefore, the total polarisation behind the {\it quarter wave plate} can
be higher than the initial polarisation. This increase is more pronounced
in the case when the photon momentum makes a small angle of
$\psi$=40\,$\mu$rad with respect to the crystal plane
(Fig.~\ref{F:stokes}b). As described in section~\ref{sec:exp_quarter} and
as shown in Fig.~\ref{F:stokes}, the final calculation takes into account
the beam divergence, in both the horizontal and vertical planes.

As seen from Fig.~\ref{F:stokes} the Si crystal with a thickness of
$L>$25cm can act as a quarter wave plate taking into account the angular
divergence of the $\gamma$-beam. For these crystal thicknesses where the
$\eta_3(0)$ component of the initial linear photon beam polarisation will
be totally transformed into the final circular component $\eta_2(L)$, only
a few percent of the photons will survive.  We defined a figure of merit
(FOM), to find a compromise between the photon beam attenuation and the
polarisation transformation efficiency in~\cite{propos}, as:
\begin{equation}
FOM = \eta_2(\ell) \sqrt {N(\ell)}.   
\label{eq:no8}
\end{equation}

\begin{figure*}[ht]
\includegraphics[scale=0.624]{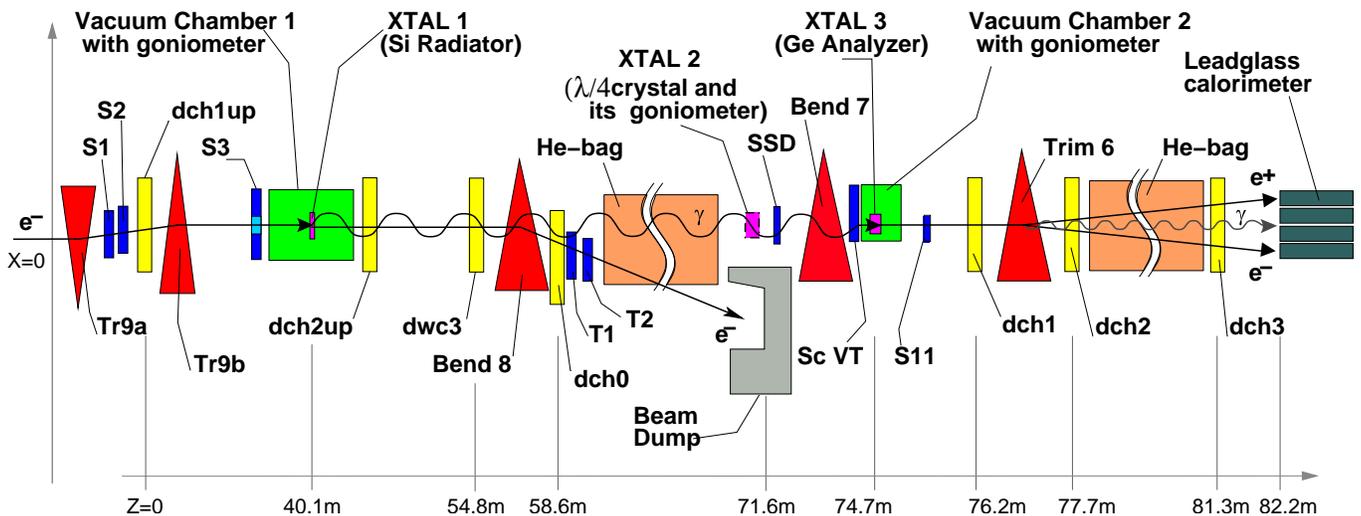}
\caption{\label{F:setup} NA59 experimental setup.}
\end{figure*}

Here $N(\ell)$ is the statistical weight of the number of surviving
photons. Taking into account equation~(\ref{eq:no8}),
references~\cite{strakh,maish2,akop} presented theoretical predictions
showing the possibility of transforming the linear polarisation of a high
energy photon beam into circular polarisation in the 70-100\,GeV energy
range. The theoretical calculations of the energy and the orientation
dependence of the indices of refraction were performed using the
quasi-classical operator method~\cite{strakh} and CPP
formulae~\cite{maish2,akop}, respectively. In these references, the
optimum thickness for a {\it quarter wave plate} Si crystal was found to
be 10\,cm. The relevant geometrical parameters involved the photon beam
forming an angle of 2.3\,mrad from the axis $(110)$ and the photon
momentum directly in the $(110)$ plane of Si single crystal, {\it i.e.}
the angle between the photon momentum and crystal plane is $\psi$=0. For
this choice of parameters, the fraction of surviving photons is 17-20$\%$.

\section{\label{apparat}Experimental apparatus}

The NA59 experiment was performed in the North Area of the CERN SPS, where
unpolarised electron beams with energies above 100\,GeV are available. We
took our data sample with 178\,GeV electrons and an angular divergence of
48\,$\mu$rad ($\sigma$) and 35\,$\mu$rad ($\sigma$) in the horizontal and
vertical planes, respectively.

The experimental setup used by NA59 is shown in Fig.~\ref{F:setup}.

The beam passes first through the 1.5\,cm thick Si crystal {\it radiator}.
This crystal is mounted in the goniometer located in the first vacuum
chamber. The electron beam makes an angle of 5\,mrad with respect to the
$<$001$>$ axis and 70\,$\mu$rad with respect to the $(100)$ plane. In this
orientation the peak intensity of the radiated photon spectrum is around
70\,GeV, and the maximum linear polarisation is $\sim$55$\%$.  Three
scintillators, S1, S2 and $\overline{\textrm S3}$ provide a coincidence
signal for the primary trigger. The drift chambers Dch1up, Dch2up and the
delay wire chamber Dwc3 define the incidence and exit angles of the
electron beam relative to the {\it radiator}.  The tagging system for the
outgoing electron consists of the dipole magnet B8, chamber Dch0, and
scintillators T1 and T2.  This tagging system will only fire if the
electron radiated at least 10\,GeV of its initial energy due to the
geometrical acceptance of scintillators T1 \& T2.

\begin{figure}
\includegraphics[scale=.44]{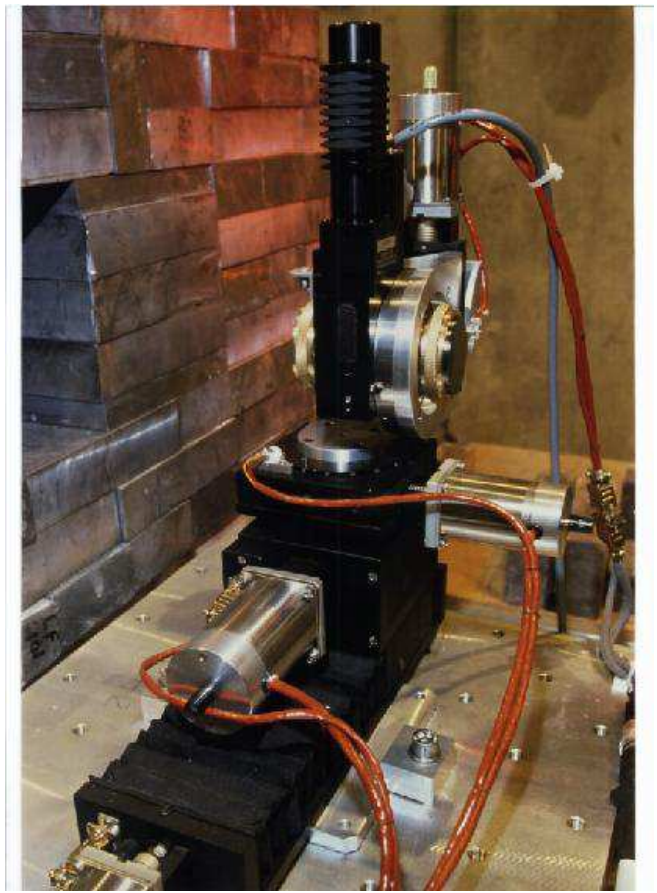}
\caption{\label{F:quarter_photo} {\it Birefringent (quarter wave plate)}
         Si crystal and goniometer.}
\end{figure}

The second goniometer needed to control the 10\,cm thick Si $<$110$>$
crystal, that served as a {\it quarter wave crystal}, is located after the
He-bag.  A photograph of the {\it quarter wave plate} and the goniometer,
at that location, is shown in Fig.~\ref{F:quarter_photo}.  The
orientation of this crystal relative to the photon beam was already
discussed in the previous section (see Table~\ref{T:crystal} for a summary
of the crystal parameters).

\begin{table*}[ht]

\begin{center}
\begin{tabular}{|c|c|c|c|c|c|}
\hline
\hline
Crystal Type   &Purpose     &Axes and Planes      &Orientation            &Thickness     \\
\hline
\hline

Si            &Radiator    &$<$001$>$, (110)      &$\theta_0$=5\,mrad, $\psi_{(110)}$=70\,$\mu$rad               &1.5\,cm      \\
\hline

Si            &Quarter Wave Plate  &$<$110$>$, (110)   &$\theta_0$=2.3\,mrad, $\psi_{(110)}$=0               &10\,cm    \\
\hline

Ge           &Analyser    &$<$110$>$, (110)   &$\theta_0$=3\,mrad, $\psi_{(110)}$=0               &1\,mm    \\
             & $\eta_1$ measurement&    &   roll wrt radiator = $\pi/4$, $3\pi/4$               &    \\
             & $\eta_3$ measurement&    &   roll wrt radiator = 0, $\pi/2$               &    \\
\hline
\hline

\end{tabular}
\caption{Where $\theta_0$ is the angle between the photon momentum and crystal axis and
$\psi$ is the angle between the photon momentum and the indicated
crystal plane.\label{T:crystal}}
\end{center}
\end{table*}

\begin{figure*}[ht]
\includegraphics[scale=0.539]{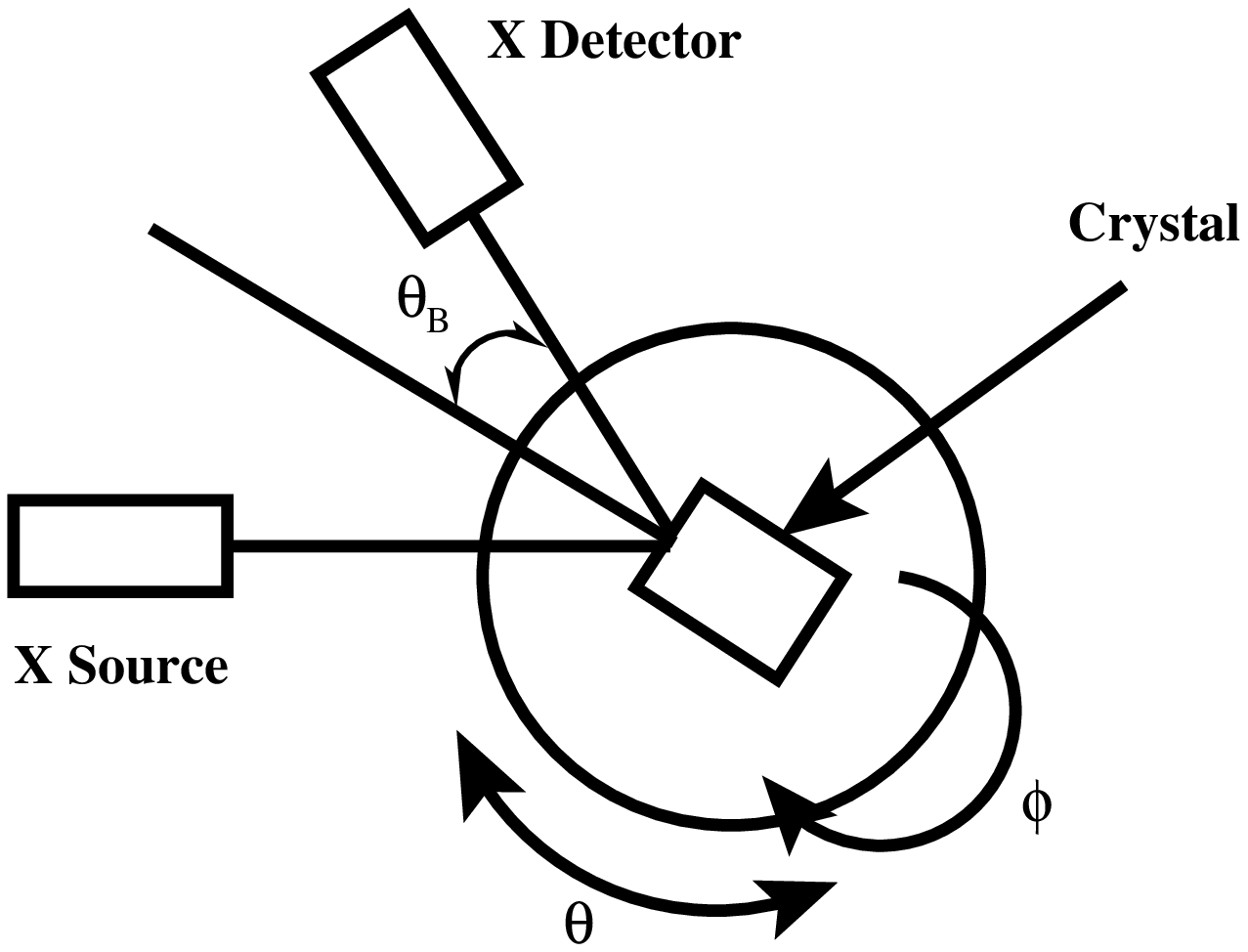}
\includegraphics[scale=0.539]{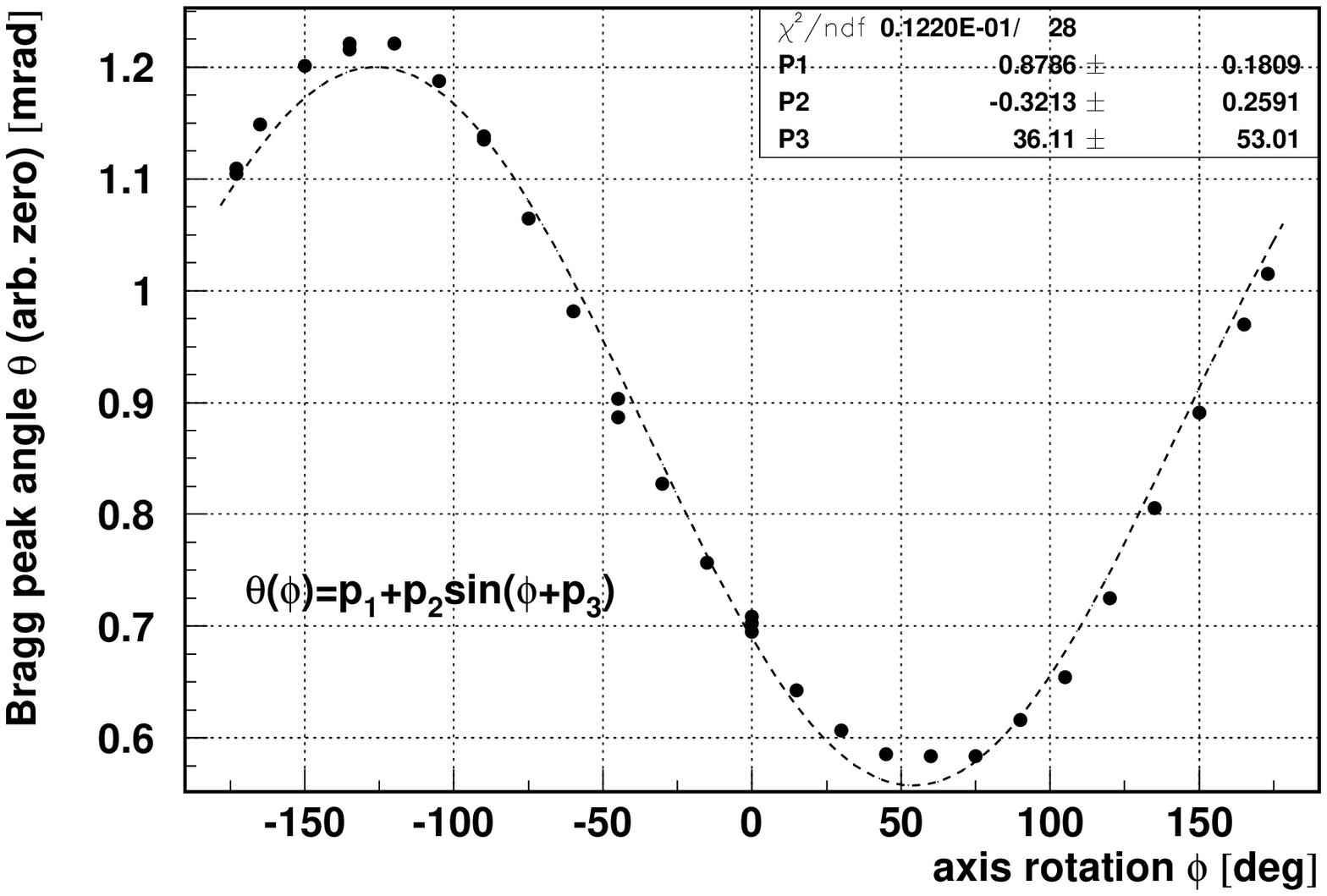}
\caption{\label{F:quarter_alig} Procedure and results for the annular 
     stage alignment used to rotate the {\it quarter wave plate} crystal 
     around the crystal axis.} 
\end{figure*}

The axis of the Si crystal was carefully pre-aligned with respect to the
axis of the azimuthal annular stage that was subsequently mounted into the
main goniometer.  This pre-alignment procedure was carried out at ESRF,
Grenoble. A schematic of the alignment setup and the results are shown in
Fig.~\ref{F:quarter_alig}. An X-ray reflection satisfying the Bragg
condition was used to monitor the orientation of the the $(110)$
crystallographic plane which is perpendicular to the $<$110$>$ axis. The
crystal was rolled in steps using the azimuthal goniometer stage ($\phi$
angle rotation). The $<$110$>$ crystallographic axis was slightly
misaligned with the crystal physical longitudinal axis and therefore also
initially slightly misaligned with the the azimuthal annular stage
longitudinal axis. At each azimuthal step the Bragg condition had
therefore to be recovered by adjustments to the angle of crystal face
using a second goniometer ($\theta$ angle rotation). The Bragg condition
was recognised by locating the two points at half-maximum of the Bragg
peak. From a plot of the adjustment angle $\theta$ for each step in the
roll angle $\phi$ of the azimuthal goniometer, the precise offset angles
between the azimuthal goniometer longitudinal axis and the $<$110$>$
crystallographic axis could be obtained. As the thick Si crystal was
mounted in the azimuthal stage by adjustment screws, the $<$110$>$
crystallographic axis could then be brought into coincidence with the
longitudinal axis of the azimuthal goniometer.

The magnet B7 served as a sweeping magnet of the particles produced by
electromagnetic showers in the {\it quarter wave plate}.

The scintillator Sc VT rejects radiation events coming from the conversion
of the tagged photon beam upstream of the crystal {\it analyser}. This
scintillator is mounted in front of the third goniometer used for the Ge
crystal. The 1\,mm thick Ge crystal served as a linear polarisation {\it
analyser} with the photon beam momentum making an angle of 3\,mrad with
respect to the $<$110$>$ axis and it also lies within the (110) plane. The
mechanisms of operation of the {\it analyser} was based on a measurement
of the PP asymmetry, for conditions where the {\it analyser} crystal had
the orientation just mentioned and/or perpendicular to it. The goniometer
allows the crystal rotation around three axes -- vertical, horizontal and
around the beam direction -- where one step is corresponds to a rotation
of 5\,$\mu$rad. The S11 scintillator was used to detect the photon
conversion into e$^+$e$^-$ pairs at the {\it analyser} and it measured the
number of charged particles seen right after the crystal {\it analyser}.
For the analysis, we only used events with 2MIPs in S11, as a signature
for PP events. The pair spectrometer consists of three drift chambers
Dch1, Dch2, Dch3 and the dipole magnet Tr6.  The trajectories of the
e$^+$e$^-$ pairs created in the {\it analyser} are tracked by the drift
chambers, and the momentum reconstruction of the pair gives us the
momentum of the incident photon. The total radiated energy is measured by
the lead glass array consisting of 13 lead glass detectors~(LG).

The various plastic scintillators, that we have already mentioned above,
were used to calibrate the tracking chambers and to define different
physics triggers. The $norm$-trigger consisted of
S1${\cdot}$S2${\cdot}\overline{\textrm S3}$ to ensure that an electron is
headed in the allowed direction of the {\it radiator} crystal. The
$rad$-trigger was defined as $norm{\cdot}$(T1.or.T2)${\cdot}\overline{VT}$
to show that the incoming electron has radiated and had been successfully
bent out of the photon beam line. The $pair$-trigger was constructed as
$rad{\cdot}$S11 to select the events for which at least one $e^{+}e^{-}$
pair was created inside the analyser crystal.

A solid state detector SSD (500\,$\mu$m thick Si crystal,
5$\times$5\,cm$^2$) was placed right after the {\it quarter wave crystal}
during dedicated runs in order to study the shower development.

Further details about the setup and the crystal alignment are given
in~\cite{na59cb}.

\section{\label{results}Results and Discussion}

The experiment was performed in two stages. In the first stage we studied
in detail the linear polarisation of the photon beam, see
Ref.~\cite{na59cb}.  In the second stage, the {\it quarter wave plate}
crystal was introduced between the {\it radiator} and {\it analyser}
crystal to investigate the transformation of the linear polarisation of
the photon beam into circular polarisation.  In this section, we describe
first the method used to measure the linear polarisation, followed by a
summary of the results presented in Ref.~\cite{na59cb}, and then present
the {\it quarter wave plate} results. The linear polarisation measurements
are important and mentioned here because the technique of identifying
circular polarisation is related to a reduction in linear polarisation and
the conservation of polarisation.

\subsection{Polarisation measurement method}

Both the $\eta_1$ and $\eta_3$ Stokes parameters described the linear
polarisation of the photon beam, and were measured using a Ge crystal as
an {\it analyser} with the method proposed in Ref.~\cite{barb}. For the
polarisation measurement, the experimentally relevant quantity is the
asymmetry, $A$, between the PP cross sections, $\sigma$, of parallel and
perpendicularly polarised photons, where the polarisation direction is
measured with respect to a particular crystallographic plane of the {\it
analyser} crystal. This asymmetry is related to the linear polarisation of
the photon, $P_{\rm l}$, through:
\begin{equation}
A  \equiv   \frac{\sigma (\gamma _{\perp }\rightarrow e^{+}e^{-})-\sigma
(\gamma _{\parallel }\rightarrow e^{+}e^{-})}{\sigma
(\gamma _{\perp }\rightarrow e^{+}e^{-})+\sigma (\gamma _{\parallel 
}\rightarrow e^{+}e^{-})}
=R \times P_{\rm l},
\label{eq:no9}
\end{equation}
where $R$ is the so called ``analysing power'' of the {\it analyser}
crystal. The quantity $R$ corresponds to the asymmetry of a 100\% linearly
polarised photon beam, and it can be computed reliably~\cite{na59cb}.

If we assume that the efficiencies and acceptances are the same between
the $pair$-trigger and the beam intensity normalisation trigger, taken
from the $rad$-trigger, then the cross section measurement in
equation~(\ref{eq:no9}) reduces to counting these events separately, and
the measured asymmetry can be written as:
\begin{equation}
A=\frac{p_{\perp}/n _{\perp} - p_{\parallel }/n_{\parallel }} 
{p _{\perp}/n _{\perp} + p_{\parallel }/n_{\parallel }},
\label{eq:no10}
\end{equation}
where $p _{\perp}$ and $p_{\parallel }$ are the number of pairs produced
in the perpendicular and parallel data configurations, while $n _{\perp}$
and $n_{\parallel }$ are the corresponding normalisation events.

\subsection{Linearly Polarised Photon Beam -- Aligned Radiator}

\begin{figure}[ht]
\includegraphics[scale=0.468]{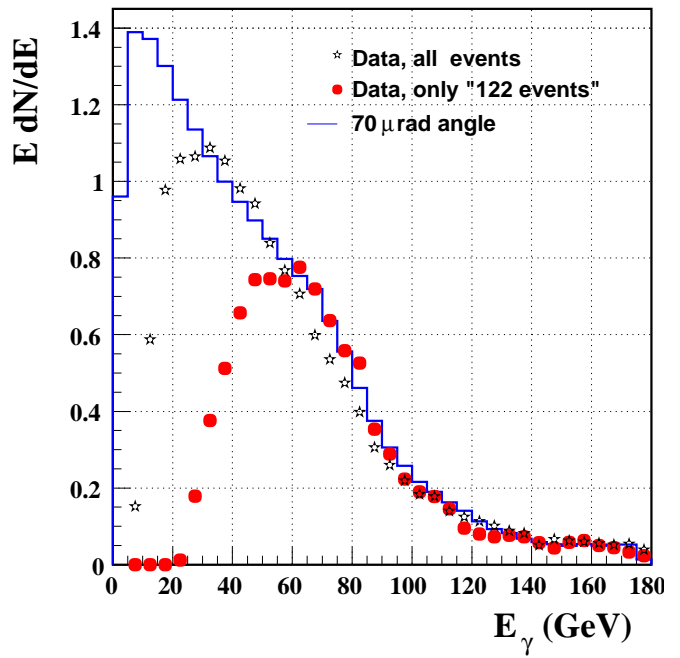}
\caption{\label{F:side:ph-int} Single photon intensity without a {\it 
    quarter wave plate}. The label ``122 events'' refers to the cleanest 
    ones with one hit in the chamber before the spectrometer magnet 
    ($e^+e^-$ pair are together), and two in both the second and third 
    downstream drift chambers ($e^+e^-$ pair are bend in opposite 
    direction by the spectrometer magnet, and therefore give two distinct 
    signals in the bending plane).}
\end{figure}   

\begin{figure}[ht]
\includegraphics[scale=0.433]{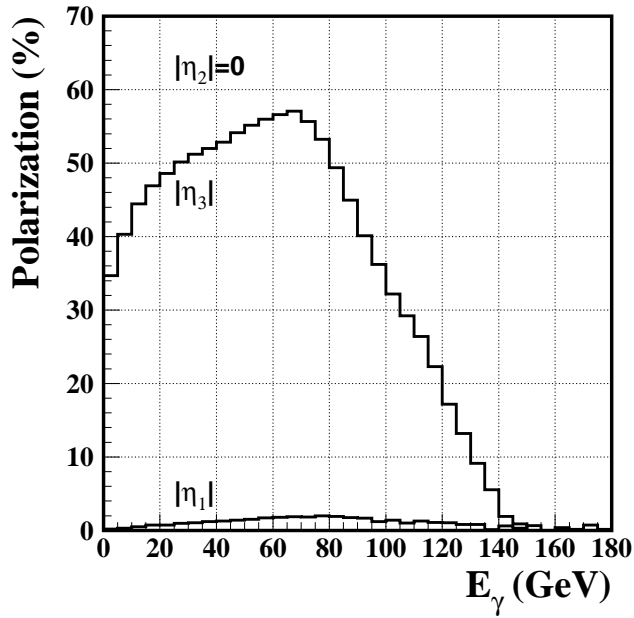}
\caption{\label{F:side:rad-pol} Stokes parameters without a {\it quarter 
          wave crystal}.}
\end{figure}

The expected and measured single photon spectrum for the chosen CB
parameters for the {\it radiators} is shown in Fig.~\ref{F:side:ph-int}.  
It is evident that there is a good agreement between them.

The expected polarisation for the same set of parameters is given in
Fig.~\ref{F:side:rad-pol} as a function of the single photon energy. As
shown in reference~\cite{na59cb}, we have confirmed from our polarisation
measurements the degree of polarisation for the $\eta_1$ and $\eta_3$
component.

\begin{figure}[ht]
\includegraphics[scale=0.558]{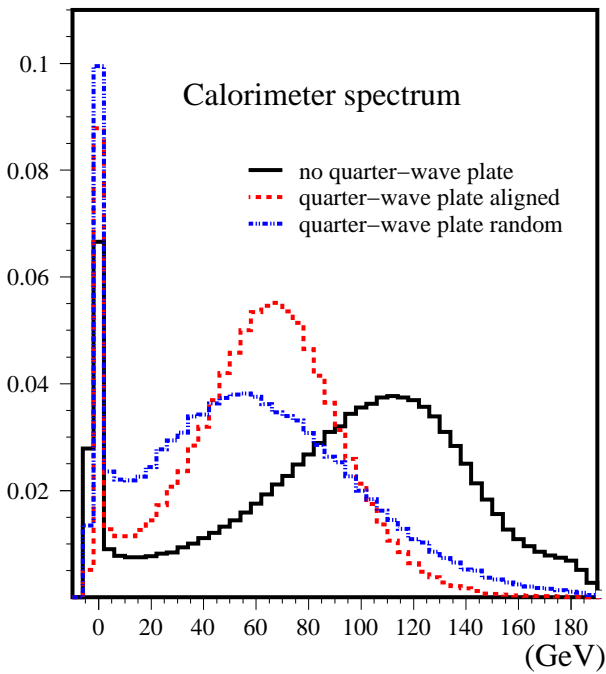}
\caption{\label{F:l4-calo} Change in measured calorimeter energy spectrum 
    for different settings of the {\it quarter wave crystal}. The shift in 
    the peak energy is clear.}
\end{figure}

The most important facts from our previous measurements are: (1) there is
a good agreement for the expected single photon spectrum, therefore any
change in the single photon spectrum after adding the {\it quarter wave
crystal} reflects how the incoming photons are absorbed or transformed by
it; (2) $\eta_1$ was found to be consistent with zero, therefore any
nonzero value observed after adding the {\it quarter wave crystal} is a
reflection of birefringent effects of the crystal.

\begin{figure}[ht]
\includegraphics[scale=0.433]{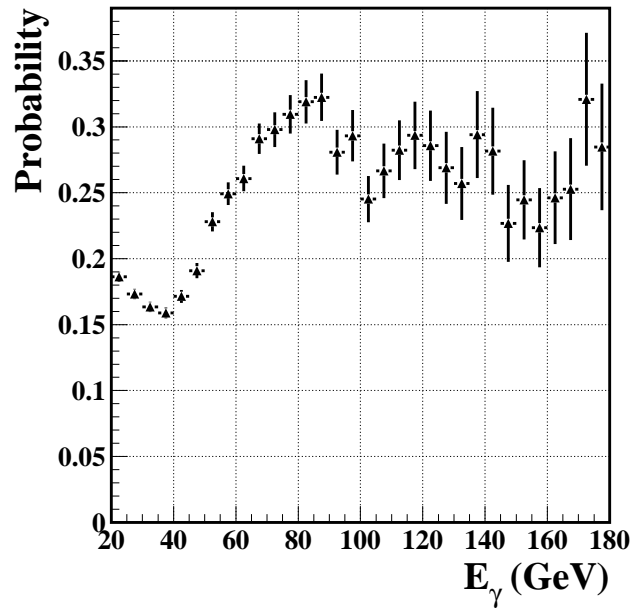}
\caption{\label{F:absor} Absorption probability found from the ratio of 
     the single photon spectrum for the data with and without the {\it 
     quarter wave crystal}, see Fig.~\ref{F:side:ph-int}.}
\end{figure}

\subsection{\label{sec:exp_quarter}Elliptically polarised photon beam -- 
Aligned Radiator \& Quarter Wave Crystal}

The birefringent properties of the oriented single crystal are
investigated by letting the linearly polarised photon beam pass through
the 10\,cm thick Si crystal. Detailed theoretical calculations and
simulations have been done to choose the crystal type, orientation and
optimal thickness. That analysis took into account the real experimental
parameters including the angular spread of the incident photon beam, the
generation of secondary particles, multiple Coulomb scattering, and all
particles produced by electromagnetic showers were also taken into
account. In the simulation we assume the angular spread of the photons
with energies between 70-100\,GeV to be about $\sim$60\,$\mu$rad and
$\sim$45\,$\mu$rad in horizontal and vertical planes, respectively, as
measured from the data. The calculations also include the polarisation
transformation part for the surviving photon beam, resulting in elliptical
polarisation.

\begin{figure*}[ht]
\includegraphics[scale=0.75]{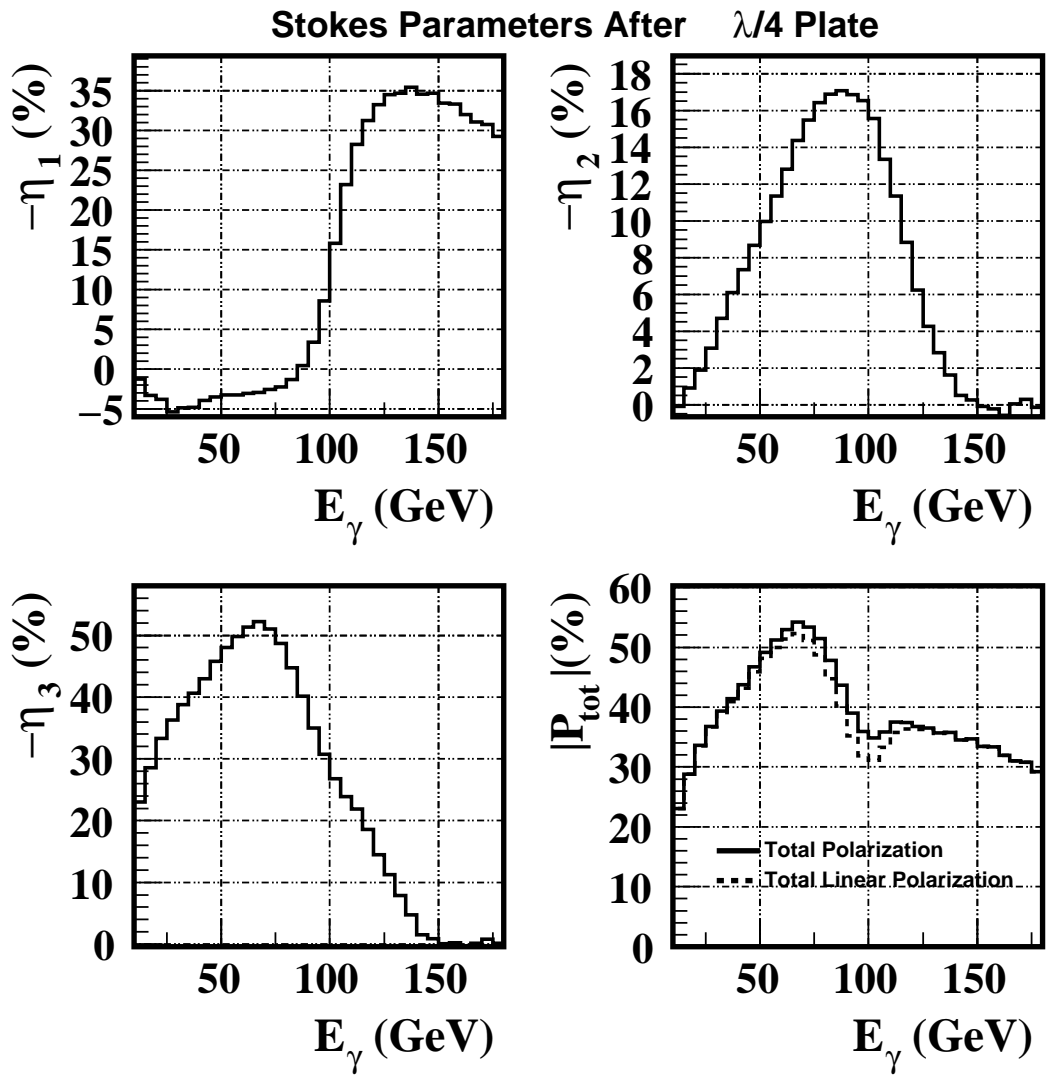}
\caption{\label{F:l4-stokes} Stokes parameters after the {\it quarter wave 
    crystal}, assuming as input the values given in 
    Fig.~\ref{F:side:rad-pol}.}
\end{figure*}

Fig.~\ref{F:l4-calo} shows the photon beam spectrum measured with the LG
electromagnetic calorimeter. The calorimeter sees all the surviving
photons radiated by the parent electron.  By comparing the spectrum with
the {\it quarter wave crystal} at random and/or aligned with the case in
which there is no {\it quarter wave crystal}, we can see that the {\it
quarter wave plate} consumes a significant amount of the beam. This causes
the peak energy of the pileup spectrum to be reduced by at least 50\,GeV.
However, it is also clear that the energy of the photons absorbed by the
{\it quarter wave crystal} depends on its alignment condition.

As already mentioned, the prediction is that only 17-20\% of the photons
will survive in our energy region.  This is confirmed by the data, see
Fig.~\ref{F:absor}.  In addition, it is clear that the survival
probability is also energy dependent as expected.

Another consequence of adding the {\it quarter wave crystal} is a
significant increase in the photon multiplicity of an event.  For example,
we expect an average multiplicity of three photons per electron for the
nominal {\it radiator} settings.  By analysing the correlation between the
calorimeter spectrum and the single photon spectrum, we can conclude that
the majority of these photons have energies $<$5~GeV, and that the
calorimeter events at high energies are dominated by a single high energy
photon and not due to the pileup of many low energy photons. As a
consequence, the measurement of the Stokes parameters in the high energy
range can be performed by measuring the asymmetry using either the
calorimeter or the pair spectrometer.

\begin{figure}[ht]
\includegraphics[scale=0.433]{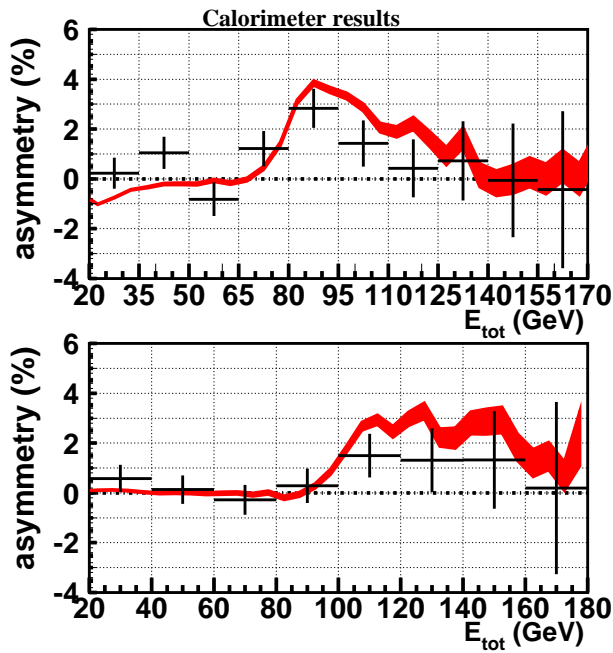}
\caption{ \label{F:l4-asy1} Asymmetry measured with the calorimeter. The 
    results reflects the changes in $\eta_3$\,(top) and generation of 
    $\eta_1$\,(bottom) due to the presence of the {\it quarter wave 
    crystal}.}
\end{figure}

\begin{figure}[ht]
\includegraphics[scale=0.473]{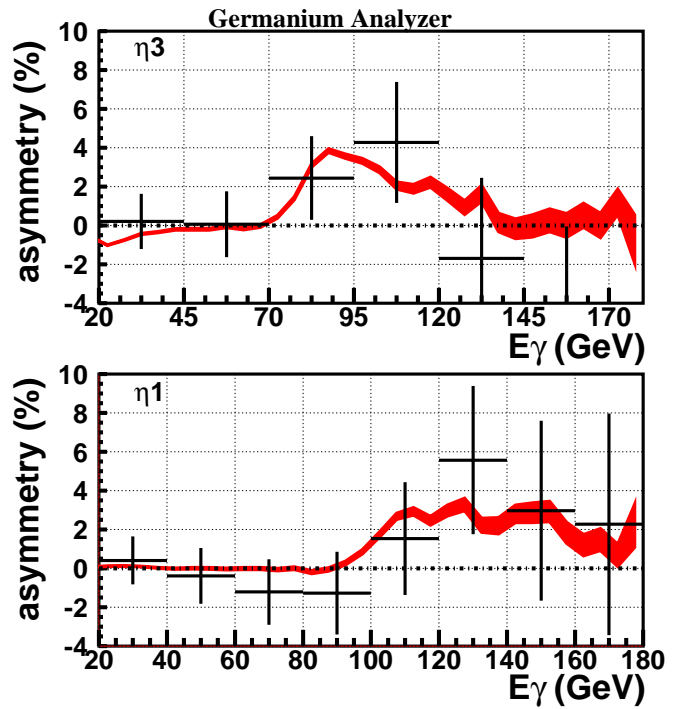}
\caption{\label{F:l4-asy2} Asymmetry measured with the pair spectrometer. 
    The data sample  with a fully reconstructed single e$^+$e$^-$ pair is 
    ten times smaller than the total data sample with at least one pair 
    passing all the data quality cuts and with a 2MIPs cut  in S11.}
\end{figure}

The expected Stokes parameters and the total polarisation of the photons
after the {\it quarter wave crystal} are given in
Fig.~\ref{F:l4-stokes}. As shown, the expected value of the $\eta_3$
Stokes parameter decreases from 36$\%$ to 30$\%$ around 100\,GeV. This
difference should be seen in the PP asymmetry. The expected degree of
circular polarisation is of the order of $\sim$16$\%$ at the same energy.
In Fig.~\ref{F:l4-stokes}, we expect an interesting increase of up to a
factor of seven for the $\eta_1$ Stokes parameter in the same energy
region.  This phenomenon was also predicted by Cabibbo~\cite{cabibbo2},
the unpolarised photon beam traversing the aligned crystal becomes
linearly polarised. This follows from the fact that the high-energy
photons are mainly affected by the PP process. This cross section depends
on the polarisation direction of the photons with respect to the plane
passing through the crystal axis and the photon momentum (polarisation
plane). Thus, the photon beam penetrating the oriented single crystal
feels the anisotropy of the medium. For the experimental verification of
this phenomenon with photon beams at energies of 9.5\,GeV and 16\,GeV,
see~\cite{berger, eisele}.  In the high energy region $>$100\,GeV the
difference between the PP cross sections parallel and perpendicular to the
polarisation plane is large.  Since the photon beam can be regarded as a
combination of two independent beams polarised parallel and perpendicular
with respect to the polarisation plane, one of the components will be
absorbed to a greater degree than the other one, and the remaining beam
becomes partially linearly polarised.

The data taking was limited to 18~hours due to a breakdown in the SPS that
lead to a six day shutdown. As a consequence, we have a small data sample
to test the predictions described above. The measured asymmetries using
the calorimeter are given in Fig.~\ref{F:l4-asy1} and again using the
pair-spectrometer in Fig.~\ref{F:l4-asy2}.  In order to reduce
systematic uncertainties, the angular settings of the {\it radiator}
crystal (hence the direction of linear polarisation of photon beam) were
kept constant, and only the {\it analyser} crystal was rolled around its
symmetry axis to obtain the parallel and perpendicular configurations.  
Therefore, to measure the polarisation of the $\eta_{3}$ ($\eta_{1}$)
component, the asymmetry between the 0 ($\pi /4$) and $\pi /2$~($3\pi /4$)
{\it analyser} orientations were used.

As shown in these figures, the measured asymmetries are in agreement with
the predicted polarisation for the chosen Ge {\it analyser} crystal
setting~\cite{na59cb}. For the Stokes parameter $\eta_{3}$, the measured
asymmetry after the {\it quarter wave crystal} is about 2.9$\pm$0.7\% in
the energy range between 80-100\,GeV. The estimated analysing power $R$
for the Ge~{\it analyser} in the same energy range is about
10.4\%~\cite{na59cb}. Using the equation~(\ref{eq:no9}) one can estimate
the measured Stokes parameter $\eta_3$ after the {\it quarter wave
crystal}. Thus, the measured Stokes parameter is $\eta_3$=28$\pm$7\% (see
Fig.~\ref{F:l4-stokes}). For the Stokes parameter $\eta_3$, the measured
asymmetry without the {\it quarter wave crystal} in the same energy range
was found to be 4.7$\pm$1.7\%, (see~\cite{na59cb}). This corresponds to a
measured value of $\eta_3$=44$\pm$11\%, which is also consistent with the
theoretically expected value of $\eta_3$, see Fig.~\ref{F:side:rad-pol}.
The experimental measured and predicted degree of linear polarization
(Stokes parameters $\eta_1$ and $\eta_3$) are presented
in~\cite{overview}.

Similar calculations may be done for the Stokes parameter $\eta_1$. If we
make a weighted average for the asymmetry values between 20 and 100 GeV,
where we expect no asymmetry, we obtain a value of 0.19$\pm$0.3\%. Above
100\,GeV we expect a small asymmetry, where we measured (1.4$\pm$0.7)\%
between 100 and 180\,GeV.

Using the equation~(\ref{eq:no7}) one can now find the measured circular
polarization degree which is equal $\eta_2$=21$\pm$11\%. This consistent
with the predicted value of 16\%.

The statistical significance of the result was estimated using the F-test
to evaluate the confidence level associated with distinguishing between
two different statistical distributions. The first distribution was formed
by the variance of the energy dependent data for the experimental circular
polarisation with respect to the theoretical prediction displayed in
Fig.~\ref{F:l4-stokes}. The second distribution was formed by the
variance of the same data to the null hypothesis prediction of no circular
polarisation. Limiting the test to the region where the crystal
polarimeter has analysing power, and also to the region where the circular
polarimetry technique of equation~\ref{eq:no7} has efficiency (80 -
110\,GeV), then we find a confidence limit of 73\% for the observation of
circular polarisation.

\section{Conclusion}

The experimental results show that the aligned single crystals are unique
tools to produce polarized high-energy photon beams and for measuring its
polarization degree. Coherence effects in single crystals can be used to
transform linear polarization of high-energy photons into circular
polarization and vice versa. Thus, it seems possible to produce circularly
polarized photon beams with energies above 100\,GeV at secondary
(unpolarized) electron beams at high energy proton accelerators.  The
birefringent effect becomes more pronounced at higher photon energy, which
allows for thinner crystals with higher transmittance. Existing diamond
arrays have already the appropriate quarter lambda thickness of about 2 cm
and will have a transmission probability of about 80\% for 100 GeV
photons. Such a diamond array that could act as a {\it quarter wave
crystal} was produced for our collaboration and was aligned and used as a
linear polarization analyser, see Ref.~\cite{na59cb}.

We did not perform a direct measurement of the circular polarization by
measuring the asymmetry of $\rho$-meson decays, which would have needed
additional beam time. However, realistic theoretical calculations describe
very well the radiated photon spectrum from the aligned radiator and the
pair production asymmetries in the aligned analyser both with and without
the birefringent Si crystal in the photon beam. In view of this good
agreement the predicted birefringent effect seems to be confirmed by the
present measurements. Measurements of the charged particle multiplicity
with depleted Si detectors show a large sensitivity to crystal alignments
and can be used to control the alignment of crystals and photon
polarization in a future polarimeter set-up.

\indent
\begin{acknowledgments}

We dedicate this work to the memory of Friedel Sellschop. We express our
gratitude to CNRS, Grenoble for the crystal alignment and Messers DeBeers
Corporation for providing the high quality synthetic diamonds.  We are
grateful for the help and support of N. Doble, K. Elsener and H. Wahl. It
is a pleasure to thank the technical staff of the participating
laboratories and universities for their efforts in the construction and
operation of the experiment.

This research was partially supported by the Illinois Consortium for
Accelerator Research, agreement number~228-1001. UIU acknowledges support
from the Danish Natural Science research council, STENO grant no
J1-00-0568.

\end{acknowledgments}

\bibliography{na59-l4}

\end{document}

%% file: na59-l4-authors.tex
\author{A.~Apyan}
\altaffiliation[Now at: ]{Northwestern University, Evanston, USA}
\affiliation{Yerevan Physics Institute, Yerevan, Armenia}

\author{R.O.~Avakian}
\affiliation{Yerevan Physics Institute, Yerevan, Armenia}

\author{B.~Badelek}
\affiliation{Uppsala University, Uppsala, Sweden}

\author{S.~Ballestrero}
\affiliation{INFN and University of Firenze, Firenze, Italy}

\author{C.~Biino}
\affiliation{INFN and University of Torino, Torino, Italy}
\affiliation{CERN, Geneva, Switzerland}

\author{I.~Birol}
\affiliation{Northwestern University, Evanston, USA}

\author{P.~Cenci}
\affiliation{INFN, Perugia, Italy}

\author{S.H.~Connell}
\affiliation{Schonland Research Centre - University of the Witwatersrand,
Johannesburg, South Africa}

\author{S.~Eichblatt}
\affiliation{Northwestern University, Evanston, USA}

\author{T.~Fonseca}
\affiliation{Northwestern University, Evanston, USA}

\author{A.~Freund}
\affiliation{ESRF, Grenoble, France}

\author{B.~Gorini}
\affiliation{CERN, Geneva, Switzerland}

\author{R.~Groess}
\affiliation{Schonland Research Centre - University of the Witwatersrand,
Johannesburg, South Africa}

\author{K.~Ispirian}
\affiliation{Yerevan Physics Institute, Yerevan, Armenia}

\author{T.J.~Ketel}
\affiliation{NIKHEF, Amsterdam, The Netherlands}

\author{Yu.V.~Kononets}
\affiliation{Kurchatov Institute, Moscow, Russia}

\author{A.~Lopez}
\affiliation{University of Santiago de Compostela, Santiago de Compostela,
Spain}

\author{A.~Mangiarotti}
\affiliation{INFN and University of Firenze, Firenze, Italy}

\author{B.~van~Rens}
\affiliation{NIKHEF, Amsterdam, The Netherlands}

\author{J.P.F.~Sellschop}
\altaffiliation[Deceased]{}
\affiliation{Schonland Research Centre - University of the Witwatersrand,
Johannesburg, South Africa}

\author{M.~Shieh}
\affiliation{Northwestern University, Evanston, USA}

\author{P.~Sona}
\affiliation{INFN and University of Firenze, Firenze, Italy}

\author{V.~Strakhovenko}
\affiliation{Institute of Nuclear Physics, Novosibirsk, Russia}

\author{E.~Uggerh{\o}j}
\thanks{Co-Spokeperson}
\affiliation{Institute for Storage Ring Facilities, University of Aarhus,
Denmark}

\author{U.I.~Uggerh{\o}j}
\affiliation{University of Aarhus, Aarhus, Denmark}

\author{G.~Unel}
\affiliation{Northwestern University, Evanston, USA}

\author{M.~Velasco}
\thanks{Co-Spokeperson}
\altaffiliation[Now at: ]{Northwestern University, Evanston, USA}
\affiliation{CERN, Geneva, Switzerland}

\author{Z.Z.~Vilakazi}
\altaffiliation[Now at: ]{University of Cape Town, Cape Town, South Africa}

\affiliation{Schonland Research Centre - University of the Witwatersrand,
Johannesburg, South Africa}

\author{O.~Wessely}
\affiliation{Uppsala University, Uppsala, Sweden}

\collaboration{The NA59 Collaboration}